# Swept-Frequency Drumhead Mechanical Resonators

Raphael St-Gelais[1,2], Simon Bernard[1], Christoph Reinhardt[1], Jack C. Sankey*[1]

[1]Department of Physics, McGill University, Montréal, Québec, H3A 2T8, Canada
[2]Department of Mechanical Engineering, University of Ottawa, Ottawa, Ontario, K1N 6N5, Canada
*email: jack.sankey@mcgill.ca

**Abstract:** We demonstrate a high-Q (>5×10⁶) swept-frequency membrane mechanical resonator achieving octave resonance tuning via an integrated heater and an unprecedented acceleration noise floor below 1 µg Hz$^{-1/2}$ for frequencies above 50 kHz. This device is compatible with established batch fabrication techniques, and its optical readout is compatible with low-coherence light sources (e.g., a monochromatic light-emitting diode). The device can also be mechanically stabilized (or driven) with the same light source via bolometric optomechanics, and we demonstrate laser cooling from room temperature to 10 K. Finally, this method of frequency tuning is well-suited to fundamental studies of mechanical dissipation; in particular, we recover the dissipation spectra of many modes, identifying material damping and coupling to substrate resonances as the dominant loss mechanisms.

The prospect of frequency tunability represents a major advantage for mechanical sensors relying on high-quality-factor (Q-factor) resonances. For example, optomechanical synchronization[1,2] ideally relies on multiple oscillators having the same frequency, which is hard to achieve due to unavoidable fabrication process variations. Frequency tuning is also of great interest for force or acceleration sensing, as it enables resonant detection that is limited by fundamental thermomechanical noise over a swept range, without the need for a precision interferometer[3]. In contrast, fixed-frequency sensors are usually limited by laser shot noise[4,5] or Johnson noise[6] at frequencies not far from the mechanical resonance.

Silicon nitride membranes are widely used for achieving high-Q mechanical resonators[7], but previous demonstrations of frequency tuning in such platform relied on complex approaches such as substrate bending[8], laser heating[9], or optical gradient forces[10]. Although successful in achieving large tuning ranges (up to 3 octaves[8]), these approaches require cumbersome external apparatus (e.g., macroscopic force transducers[8], high power lasers[9], tunable lasers[10]) that are difficult to scale to widespread practical applications. Simple electrical control would be ideal, but was reported only in emerging material platforms, such as graphene[11] or carbon nanotubes[12].

Here we demonstrate octave frequency tuning of a high Q-factor SiN drum resonator using a simple integrated electrical heater. The device (Fig 1(a)), consist of a $l$ = 1.05 mm wide, $h$ = 100 nm thick, low-stress SiN membrane—similar to membranes commonly used in high-finesse optomechanical interferometers[7]—upon which a 50-nm-thick platinum (Pt) "labyrinth" resistor is deposited (see Methods). As we apply heating current, thermal expansion reduces the membrane's tensile stress and, consequently, the mechanical resonance frequency ($f$) as:

$$\frac{f}{f_0} = \sqrt{1 - \frac{E_1 \alpha}{\sigma_0} \Delta T}, \quad (1)$$

where $f_0$ is the frequency under no heating power, $E_1 \approx$ 300 GPa is the Young's Modulus of SiN, $\sigma_0 =$ 230 MPa is the membrane's tensile stress at room temperature (calculated below), $\Delta T$ is the temperature increase imposed by the heater, and $\alpha$ is the membrane's coefficient of thermal expansion.

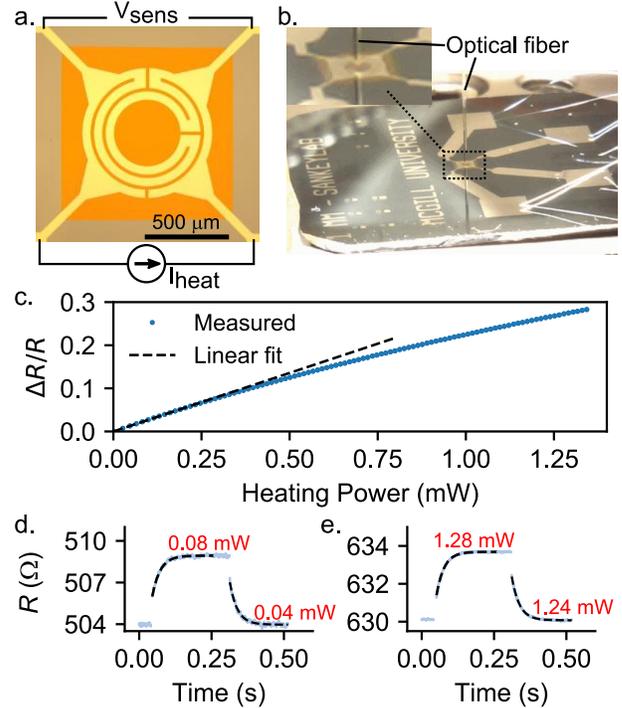

**Figure 1**: (a) SiN membrane with integrated Pt heaters, and schematic of the electrical heating circuit. (b) View of the membrane inside a high-vacuum characterization setup. (c) Dependence of labyrinth resistance on heating power. (d, e) Step response of resistance (blue lines) at low (d) and high (e) temperatures. Dashed lines are exponential fits from which we extract the thermal response time $\tau = 27.0 \pm 0.5$ ms.

**Thermal response characterization**

We first characterize the thermal response of the membrane to relate $\Delta T$ in Eq. (1), to the applied heating power $P_{heat} = I_{heat} \times V_{sens}$ (for heater current $I_{heat}$ and sensed voltage $V_{sens}$ defined in Fig. 1(a)). We measure the variation of the heater resistance ($\Delta R/R$) as a function of $P_{heat}$, which we expect to follow:

$$\frac{\Delta R}{R} = \frac{\beta \, P_{heat}}{\kappa}, \quad (2)$$

to leading order, where $\beta$ is the temperature coefficient of resistance of Pt (K$^{-1}$), and $\kappa$ is the heat conduction (W/K) between the heater and the thermal background (i.e., the silicon frame). All thermal measurements are carried out under high vacuum ($\lesssim 10^{-5}$ Torr). As shown in Fig. 1(b), we observe a linear response up to approximately 0.25 mW, at which point $\Delta R/R$ varies nonlinearly with $P_{heat}$. To determine whether this is caused by $\beta$ or $\kappa$, we measure the membrane's thermal response time $\tau = C \, m_{th}/\kappa$ (where $C$ is the heat capacity in J kg$^{-1}$K$^{-1}$, and $m_{th}$ is the thermal mass of the membrane in kg), which only depends on $\kappa$. We find that $\tau$ changes by less than our 2% measurement uncertainty between low temperature (Fig. 1(d), $\tau = 27.0 \pm 0.5$ ms) and high temperature (Fig. 1(e), $\tau = 27.1 \pm 0.2$ ms), indicating that $\kappa$ is essentially fixed and $\beta$ is the dominant source of nonlinearity. A fixed heat conduction ($\kappa$) imposes that the steady state temperature ($\Delta T$) varies linearly with the applied power as:

$$\Delta T = \frac{P_{heat}}{\kappa}, \quad (3)$$

where $\kappa = 6.1 \pm 0.5 \, \mu$W/K is calculated using the fitted (low power) slope in Fig. 1(c) and Eq. (2), with $\beta = 0.00170 \pm 0.00015 \, \text{K}^{-1}$ inferred[13] from the measured resistivity of our deposited Pt film (45 μΩ cm); this value is also consistent with direct $\beta$ measurements on similar Pt heaters[14]. With this $\kappa$ value, the highest heating power used in all reported measurements (~ 1.3 mW, limited by heater degradation) corresponds to a membrane temperature change $\Delta T = 210$ K.

**Frequency tuning and Q-factor measurements**

Such a temperature change should enable more than an octave of frequency tuning according to Eq. (1). To verify this, we measure the frequency of multiple mechanical modes as a function of temperature by monitoring the spectrum of displacement noise arising from optical interference between the membrane and the tip of a cleaved optical fiber using 1550-nm light (see, e.g., Ref 15). Measurements are carried out under high vacuum ($\lesssim 10^{-5}$ Torr) to prevent viscous damping. The optical fiber is initially positioned near the center of the labyrinth where no metal is present (Fig. 1(a)) to avoid optical heating, and offset 150 μm toward one corner to enhance the readout of membrane modes having a node at the center of the membrane (i.e., modes having even $m$, $n$ indices, where $m$ and $n$ denote the number of antinodes along the two transverse dimensions). Mechanical resonances driven by thermal noise appear as peaks in each displacement spectrum, allowing us to construct the frequency map shown in Fig. 2(a). We observe that most modes follow trajectories closely matching Eq. (1). Our simple model indeed matches well with, e.g., the fundamental (1, 1) mode trajectory when using $\alpha = 2.9 \times 10$-6 K$^{-1}$, which lies between tabulated values for SiN[16] (~2.2×10$^{-6}$ K$^{-1}$) and Pt[17] (~7.5×10$^{-6}$ K$^{-1}$) and closer to that of SiN, as expected from the relative material fractions. We observe frequency tuning over more than 1 octave, e.g., from 121.1 kHz down to 56.1 kHz for the fundamental mode.

Of central interest, we observe that the mechanical Q-factors remain high throughout their frequency tuning ranges. In Fig. 2(b) we present the first 3 symmetric modes (i.e., $m = n$, see Fig. 2(a)), along with three "typical" higher-order modes chosen to encompass more frequencies. For each mode, we measure the Q-factor at multiple heating powers by exciting a ~1 nm amplitude resonance with a piezoelectric actuator, and then cutting the drive to measure a mechanical ringdown. When no heating current is applied, all modes but one have mechanical Q-factors greater than a million, with the (3, 3) mode achieving $Q = 5.3 \times 10$-6. Strikingly, such Q-factors are comparable to values reported for metal-free SiN membranes[18,19], and for metalized SiN membranes having no metal crossing the membrane anchoring points on the silicon frame[19]. Reducing the tensile stress (via the heater) leads to a systematic reduction of the Q-factor which, as discussed in greater details below, is consistent with material damping in the metal film dominating dissipation. This systematic reduction reaches at most a factor 5 over the full tuning range, with the exception of sharp dips at frequencies matching those of the supporting frame resonances; these are also discussed in more detail below.

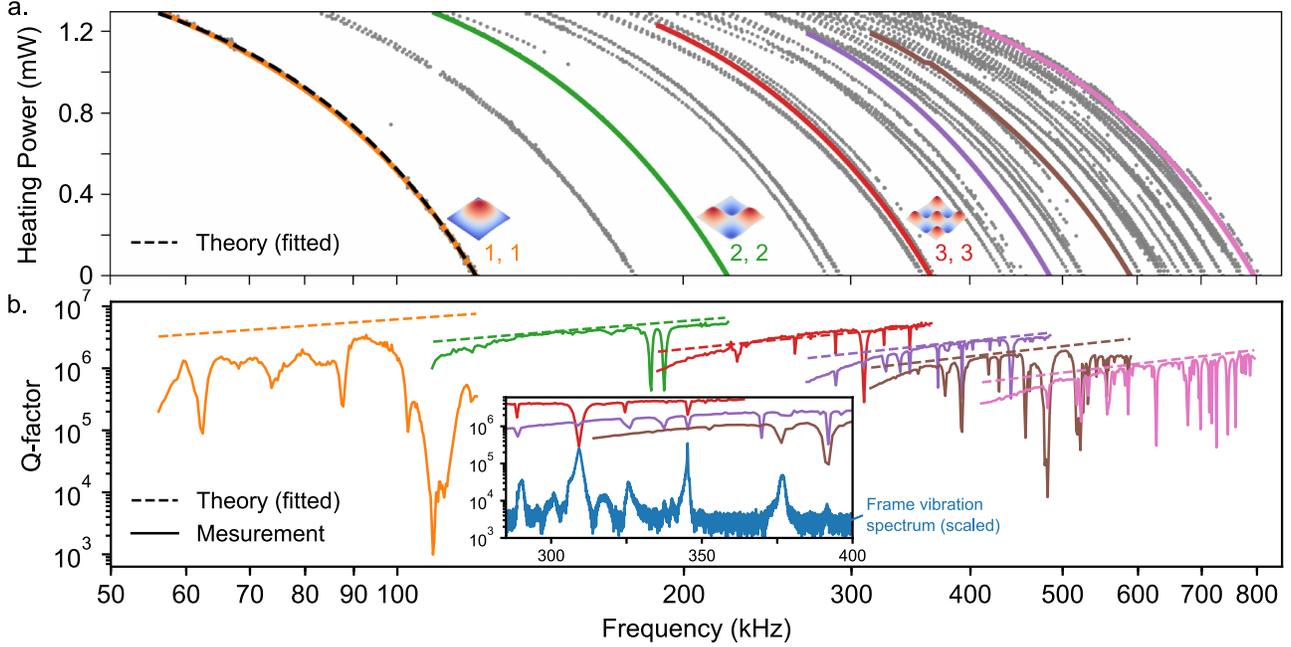

**Figure 2**: (a) Measured frequency tuning of several mechanical modes as a function of applied heating power. Colored lines correspond to modes for which Q-factors measurements are also performed. (b) Measured and predicted Q-factors for six individual mechanical modes as a function of frequency (set by heating power). Inset: comparison of Q-factor spectra with measured (piezo-driven) vibration modes of the silicon frame (scaled vertically for clarity). Colors are consistent throughout the figure, such that panel (a) can be used to assess the heating power corresponding to any given frequency in panel (b).

**Origins of mechanical dissipation**

This frequency tuning approach also provides a testbed for understanding (and ultimately minimizing) mechanical losses in membrane resonators. To understand the general trend in Q-factor, we first rule out viscous damping and acoustic radiation as possible dominant dissipation mechanisms in our system. From the geometry of our membrane and our operating pressure ($\lesssim 10^{-5}$ torr), we expect a viscous damping limit[20] $Q_{visc} \gtrsim 2 \times 10^7$ in the worst case scenario (i.e., for $f = 60$ kHz) and $Q_{visc} \gtrsim 10^8$ at the frequency where we measure the highest Q-factor (i.e., $f = 360$ kHz). Similarly, we expect the acoustic loss[21] limit for our membrane to reach $Q_{rad} \gtrsim 4 \times 10^8$ (calculated for $m, n = 1, 1$, using the expression provided in Ref. 22).

Having ruled out these contributions, we expect our system to be limited by material damping, and we confirm this trend by comparing our results with the structural damping model of Ref. 19. According to this model, the Q-factor should follow:

$$Q = \frac{E_1}{\lambda}\left(E_{2,edge} + \frac{E_{2,center}\lambda(m^2 + n^2)\pi^2}{4}\right)^{-1}, \quad (4)$$

where $E_1$ and $E_2$ are the real and imaginary part of the membrane Young's modulus, $\lambda = \sqrt{\frac{E_1 h^2}{3\sigma l^2(1-\nu^2)}}$ is a dimensionless stress parameter that depends on temperature through $\sigma = \sigma_0 - E_1 \alpha \Delta T$, and $\nu = 0.27$ is Poisson's ratio. This model accounts for structural damping at the edges (first term in Eq. (4)) and at the center of the membrane (second term). We use different loss moduli $E_{2,edge}$ and $E_{2,center}$ for the edge and center regions of the membrane (respectively) to account for different metal fractions in the two regions (see Fig. 1(a)).

To compare Eq. (4) with our results, we first extract the tensile stress $\sigma_0$ from our data using the observed resonance frequency $f_{mn}$ which, for $\Delta T = 0$, should follow[19]:

$$f_{mn} = \frac{1}{4}\sqrt{\frac{\sigma_0(m^2 + n^2)h}{m_{eff}}}, \quad (5)$$

where $m$ and $n$ are the membrane mode indices and $h = 100$ nm is the membrane thickness. We estimate the effective mass $m_{eff} = 200$ ng for the fundamental mode by assuming a sinusoidal displacement profile with a position-dependent material density $\rho$ to account for the presence or absence of metal (assuming[13] $\rho_{SiN} = 2.9$ g/cm³ and $\rho_{Pt} = 17$ g/cm³). Using this $m_{eff}$ value and the measured fundamental mode frequency $f_{11} = 121.1$ kHz in Eq. (5), we extract $\sigma_0 = 230$ MPa at room temperature. Note that this stress value is higher than in non-metalized membranes ($\sigma_0 = 130$ MPa) fabricated simultaneously on the same substrate, indicating that the Pt film is under tensile stress after deposition, consistent with previous studies of stress in sputtered metal films[23]. In Eq. (4), we also use $m^2 + n^2 \approx 2(f/f_{11})^2$ for the three higher frequency modes in Fig. 2.

We find that Eq. (4) captures the observed trends over the full data set for $E_{2,center} = 0.015$ GPa and $E_{2,edge} = 0.0015$ GPa. The fact that the loss modulus is much lower at the edges than at the center suggests that confining the Pt to the membrane corners (where the bending is not as sharp) and minimizing their widths is an effective strategy for reducing structural damping at the clamping points[19]. At the highest temperatures, measured $Q$ values consistently deviate from the model, likely due to variations in $E_2$ at high temperature (where the relative stresses of in Pt and SiN begin to change significantly).

The sharp dips in Q-factor present in Fig. 2(b) tend to coincide with the observable (lossier) driven silicon frame resonances (inset), as expected for degenerate hybridization between these modes[9,20]. We note also that a dip at a given frequency may or may not appear in different membrane mode frequency scans (e.g. see the 300 – 400 kHz region in Fig 2(b)), suggesting a varied level of spatial overlap between the modes (in principle allowing one to spatially decompose the silicon frame modes). This method of probing nearby sources of radiation loss presents a tool for designing frame geometries to minimize their negative effects.

**Acceleration and force sensing**

We also confirm that, despite the >200 K temperature change, this tuning method does not introduce any unexpected displacement noise in the membrane, a point of key importance for sensing applications. Figure 3(a) shows a plot of the integrated displacement noise (calibrated as in previous work[15]) for the (1, 1) and (2, 2) modes at multiple heating powers. If limited by thermo-mechanical noise, the equipartition theorem imposes a RMS displacement noise:

$$x_{rms} = \frac{1}{2\pi f}\sqrt{\frac{k_b T}{m_{eff}}}, \quad (6)$$

the range of which is plotted in Fig. 3(a). The 25% confidence intervals account for material constant uncertainties (i.e., densities, thicknesses, temperature coefficient of resistance, etc.), and a 10% measurement error that results from approximating the fiber-membrane optical interference signal by a sinusoid[15,24]. We calculate $m_{eff} \approx 240$ ng for the (2, 2) mode using the same first order approximation as for (1, 1). As shown in Fig. 3(a), we find that measured and predicted values of $x_{rms}$ agree within systematic uncertainties for both modes, indicating that the electrical heating method does not induce unexpected displacement noise.

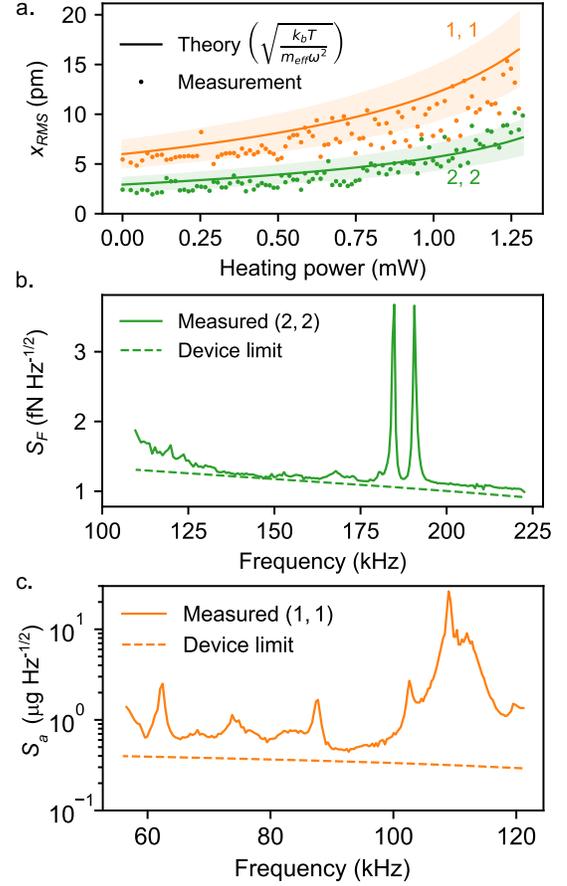

**Figure 3:** (a) Measured displacement noise as a function of heating power for two modes. Each data point is averaged for 60 sec. (b, c) Performances of the membrane as a force sensor (b) and as an accelerometer (c). Device limits are calculated from the theoretical Q-factor shown in Fig. 2(b).

Knowing that mechanical fluctuations of our membrane are consistent with fundamental thermo-mechanical noise, we now estimate the noise floor for force and acceleration sensing. Thermo-mechanical force noise is given by[25]:

$$S_F = \sqrt{\frac{8\pi k_b T f m_{eff}}{Q}}. \quad (7)$$

In turn, normalizing $S_F$ by $m_{eff}$ yields the noise equivalent acceleration[5] ($S_a$):

$$S_a = \sqrt{\frac{8\pi k_b T f}{Q m_{eff}}}. \quad (8)$$

As shown in Fig. 3(b), the (2, 2) mode achieves $S_F \sim 1$ fN Hz$^{-1/2}$, which is significantly larger than state-of-the-art[26]. We note however that our frequency tuning approach could most likely be applied to patterned membranes, such as trampolines[15,27], of much lower $m_{eff}$ in order to reach lower $S_F$.

For acceleration sensing, however, large $m_{eff}$ represents an advantage (see Eq. (8)). As shown in Fig. 3(c), the (1, 1) mode achieves a sub-µg Hz$^{-1/2}$ noise floor over the majority of the tuning range. This is an order of magnitude lower than state-of-the-art (~10 kHz) chip-scale elements[5], but the present device operates up to 120 kHz. Increasing the size and mass of our device should further reduce this noise floor (which we expect to scale as $S_a \sim f$ assuming a roughly constant $Q \times f$ product in Eq. (8)), in principle enabling low-frequency accelerometers competitive with even more massive, high-finesse interferometers[28] (our current device is already within a factor of 10).

**Bolometric optomechanical cooling**

Finally, due to its very high Q-factor and the strong mismatch of thermal expansion coefficients between Pt and SiN, our device is well-suited to low-power bolometric optomechanical actuation[29]. To demonstrate this, we position the optical fiber over the metallic region 280 µm diagonally from the membrane center (Fig. 1(a)). We estimate ~5% light absorption by comparing the membrane frequency shift to the incident laser power. Light absorption causes differential thermal stresses in the membrane, which can damp (cool) or anti-damp (drive) the membrane motion depending on the interference condition[29]. In Fig. 4, we present cooling of the (2, 2) mode measured using two separate techniques: (1) we directly measure the displacement noise and extract $T$ using Eq. (6), and (2) we measure the mechanical ringdown time, extracting the temperature $T$ using[29]:

$$\frac{T}{T_0} = \left(\frac{f_{nm}}{f}\right)^2 \frac{Q}{Q_{nm}}, \qquad (9)$$

where $f_{nm}$, $Q_{nm}$, and $T_0$ are respectively the frequency, Q-factor, and temperature in the absence of optomechanics. We performed this experiment at 0.6 mW electrical heating, thereby placing the membrane at the middle of its tuning range (and to check that heating does not interfere with bolometric optomechanics in any surprising way). Similar results are found for zero heating. As shown in Fig. 4, the measurement techniques agree with one another, demonstrating cooling from 400 K to 10 K. We find that cooling is eventually limited by instabilities at higher laser power, consistent with bolometric excitation of higher order modes[30]. Interestingly, we achieve a level of cooling comparable to previous optomechanical accelerometers[5], but we do so without a high-finesse optical resonator. We emphasize that this low-finesse operation significantly relaxes alignment, stabilization, and light source requirements. In principle, a monochromatic LED could achieve these same results, provided the coherence length (routinely several microns without a filter) is larger than the interferometer gap.

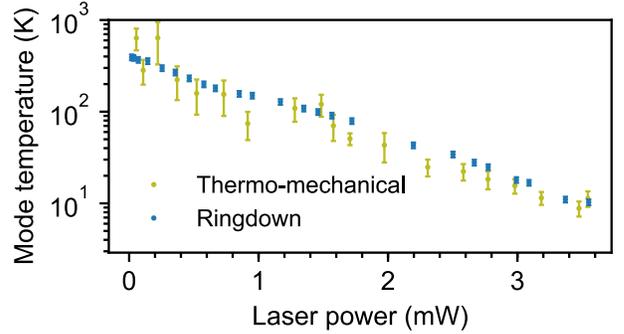

**Figure 4:** Bolometric laser cooling of the $(m, n) = (2, 2)$ mode measured using two different techniques (see main text). Error bars account for statistical measurement repeatability.

While laser cooling does not provide a fundamental improvement in noise performance (cold damping does not change coupling to the thermal bath) or bandwidth (optimal post-processing filters in principle retrieve the same information[31]), it does stabilize the membrane and reduce its response time, both of which are helpful in practical situations[5]. Laser cooling also helps in reducing the dynamic range required to resolve the full spectrum of Brownian motion. Finally, driving mechanical oscillations with the readout light source alleviates the need to incorporate a piezo driver.

**Conclusion**

We demonstrate a simple, swept-frequency accelerometer having an unprecedented noise floor over a wide range of ultrasonic frequencies. It is compatible with simple, established batch-fabrication techniques, it requires no high-quality optics or lasers to operate, and is efficiently actuated (stabilized) via bolometric optomechanics. In future work, it would be interesting to engineer a higher tuning range using heaters that can sustain higher temperatures and patterned membrane resonators such as trampolines[15,27], whose frequency depends strongly on the tension in the tethers. These devices also present an opportunity to test the spectrum of dissipation using multiple mechanical modes for each frequency in a single device.

**Methods**

Fabrication is carried out on commercially available 4 inch diameter silicon substrates coated with ~ 100 nm LPCVD (low pressure chemical vapor deposition) low-stress silicon nitride (SiN) films. We first proceed with deposition of platinum (Pt) heaters by metal sputtering (Argon gas, ~15 mTorr pressure) followed by lift-off. A ~7-nm-thick layer of chromium is used as an adhesion layer for the ~50-nm-thick Pt film. The heaters are then protected by a bilayer film of Plasma Enhanced CVD (PECVD) $SiO_2$ followed by Protek® PSB Alkaline-Protective coating. The latter protects the heaters during the substrate removal step in potassium hydroxide (KOH), while PECVD $SiO_2$ is etched in hydrofluoric (HF) at the end of the process to eliminate potential Protek residues. After the protective bilayer deposition, the substrate backside is patterned by reactive ion etching and structural release is carried out in a heated (KOH) bath. This step creates the suspended membranes and simultaneously separates the multiple dies present on the substrate. When the KOH etch step is completed, all dies are kept wet through a succession of cleaning baths. These baths consist of Nanostrip for Protek removal, deionized (DI) water rinse, 10% HF for $SiO_2$ removal, DI water rinse, and finally solvent (acetone followed by isopropanol) baths for organic residues removal. Dies are finally dried manually, out of the isopropanol bath, with a nitrogen blow gun. During the experiment heating current is applied on the membrane using a Keithley model 2400 sourcemeter.